\documentclass[mathleft]{an}

\usepackage{graphicx}
\usepackage{times}
\usepackage{natbib}
\bibpunct{(}{)}{;}{a}{}{,}
\def\note #1]{{\bf #1]}}

\begin{document}

\Pagespan{789}{}
\Yearpublication{2006}%
\Yearsubmission{2005}%
\Month{11}%
\Volume{999}%
\Issue{88}%

\title{The {\it Kepler} Asteroseismic Investigation: \\
Scientific goals and the first results}

\author{H. Kjeldsen\inst{1}\fnmsep
\thanks{\email{hans@phys.au.dk}\newline}
\and J. Christensen-Dalsgaard\inst{1}
\and R. Handberg\inst{1}
\and T. M. Brown\inst{2} 
\and
R. L. Gilliland\inst{3}
\and \\
W. J. Borucki\inst{4}
\and
D. Koch\inst{4}}
\titlerunning{Asteroseismology with {\it Kepler}}
\authorrunning{Kjeldsen et al.}
\institute{
Department of Physics and Astronomy, Building 1520, Aarhus University,
8000 Aarhus C, Denmark
\and
Las Cumbres Observatory Global Telescope, Goleta, CA 93117, USA
\and
Space Telescope Science Institute, 3700 San Martin Drive,
Baltimore, MD 21218, USA
\and
NASA Ames Research Center, MS 244-30, Moffett Field,
CA 94035, USA}

\received{????}
\accepted{????}
\publonline{later}

\keywords{stars: evolution -- stars: interiors -- stars: oscillations --
stars: planetary systems -- space vehicles}

\abstract{{\it Kepler} is a NASA mission designed to detect exoplanets
and characterize the properties of exoplanetary systems. {\it Kepler}
also includes an asteroseismic programme which is being conducted through
the Kepler Asteroseismic Science Consortium (KASC), whose
400 members are organized into 13 working groups by type of
variable star. So far data have been available from the first 7 month of the mission containing
a total of 2937 targets observed at a 1-min. cadence for periods between 10 days and 7 months. The goals 
of the asteroseismic part of the {\it Kepler} project is to perform detailed
studies of stellar interiors. The first results of the asteroseismic 
analysis 
are orders of magnitude better than seen before, and
this bodes well for how the future analysis of Kepler data for many types of
stars will impact our general understanding of stellar structure
and evolution.
}

\maketitle

\section{Introduction}

%
{\it Kepler} is a NASA Discovery mission with the primary goal to
investigate the properties of extrasolar planetary systems (exoplanets).
Of particular importance is the characterization of habitable systems, i.e.,
rocky planets, up to a few times the mass of the Earth, in a suitable distance
from the central star.
The mission detects exoplanets with the {\it transit technique}: using
a 95\,cm Schmidt telescope a 100 square degree field in Lyra and Cygnus is
observed continuously to look for small reductions in observed stellar
luminosity when a planet passes through the line of sight between the star
and the observer.
In practice, more than 150,000 stars are observed to ensure a reasonable
detection probability.
From repeated detections of such minute dimmings of the light, 
and additional follow-up observations, 
it may be confirmed that the signal is due to a planet in orbit
around the star and the properties of the planet, in particular its diameter,
can be determined \citep[see][for further details]{Boruck2010}.
A planet of the size of the Earth in orbit around a star like the 
Sun gives rise to transits with a relative decrease in the stellar flux of
around $10^{-4}$.
Such an effect is readily detectable with {\it Kepler}
\citep{Boruck2009}.

The requirement of exquisite photometric precision for the transit observations
also makes the mission perfectly suited for asteroseismology \citep{Christ2008}.
In fact, the potential for combining exoplanet search and asteroseismology
has already been demonstrated by the CoRoT mission \citep{Michel2008}.
Thus the mission has established the Kepler Asteroseismic Investigation (KAI),
with two main goals: the direct use of asteroseismology to characterize
central stars of planetary systems, and in particular determine the radius
and age of the stars;
and the use of high-quality data on a very large and diverse sample of stars
to study stellar properties and improve our understanding of stellar 
interior physics, evolution and oscillations.

To make full use of the possibilities offered by the KAI,
we have set up the Kepler Asteroseismic Science
Consortium, with direct access to the Kepler asteroseismic data.
The benefits of this organization were demonstrated by the very efficient
use of the initial data, allowing rapid publication of results of the
first month of {\it Kepler} data \citep{Gillil2010a} 
\citep[see also][for an example of the schedule of the analysis]{Karoff2010}.
Here we briefly describe the Kepler mission and the organization of the work
within KASC, as well as the parallel efforts to utilize {\it Kepler}
asteroseismology in support of the exoplanet research.

\section{The {\it Kepler} mission}

\begin{figure}
\includegraphics[width=80mm]{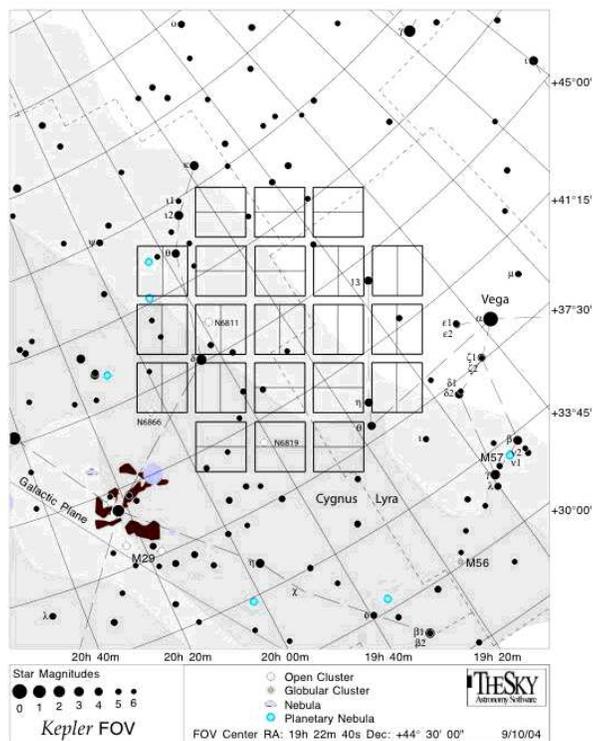}
\caption{The Kepler FOV and location of the CCDs in the sky}
\label{fig:field}
\end{figure}

%
Details on the design of the {\it Kepler} instrumentation and the operations
were provided by \citet{Koch2010}.
The {\it Kepler} photometer is of Schmidt design, with a corrector with 
a 95\,cm aperture and a 1.4\,m primary mirror.
The curved focal plane contains 42 CCD detectors, each with a
field-flattening lense.
The active field on the sky is 105 square degrees;
the field is placed such that the brightest stars, which could otherwise
cause trailing problems, are placed in the gaps between the CCDs.
The field and the location of the CCDs are shown in Fig.~\ref{fig:field}.

{\it Kepler} was launched on 7 March 2009 into a heliocentric Earth-trailing
orbit, with a period of 372.5 days.
Commissioning was carried out until 11 May 2009
and science operations started on 12 May 2009.
To keep the solar panels pointed towards the Sun the satellite is rotated
by $90^\circ$ every 93 days.
The detector plane is four-fold symmetric such that the CCDs always cover the
same areas on the sky.
Data downlink requires repointing of the satellite;
this takes place at 32-day intervals and gives rise to 24\,h 
interruptions in the observations. The nominal duration of the mission is 3.5 years.
This may be extended; the maximum duration of the mission,
limited by onboard consumables and telemetry limitations with increasing
distance is about 7-8 years.

The photometric data are downlinked in the form of small images around 
each target star, the size depending on the brightness of the star.
For stars brighter than V magnitude 11.7 the central pixels in the 
image are saturated and a larger area is required to capture the overflowing
photo-electrons;
given this, however, accurate photometry is possible for stars as bright as
magnitude 6 \citep{Gillil2010b}, 
although at the expense of telemetry.
For most targets the data are integrated over around 30 minutes, in
what is known as Long Cadence (LC) mode; this is adequate for detecting
planetary transits.
However, for up to 512 targets the data can be transmitted at a 1-minute
cadence (Short Cadence; SC);
this is required for the study of pulsations of main-sequence stars and is
also used further to characterize the planetary transits.
Of the 512 SC slots, at least 140 are reserved for targets
selected by the KASC for asteroseismology;
additional 100 slots can be used for asteroseismic targets
selected for their exoplanet interest by the Kepler Science Team.
In addition, KASC is guaranteed the right to select 1700 LC slots
for asteroseismology of more slowly pulsating stars, as well as 
access to LC data on 1000 red giants selected as astrometric references. 

Target lists are uploaded to {\it Kepler} once a quarter.
Each target (short cadence) is observed for a minimum of 1 month, although 
after the initial 10 month of operations most often a target is kept for the full quarter.
The target lists have to be ready 8 weeks before the next quarter begins.
This gives very substantial flexibility in the choice of targets for
asteroseismology, 
including taking into account results obtained in earlier phases of the 
mission, to select the optimum targets for long-term observations.
It will often be the case that a given target is observed in
several consecutive quarters.

The Kepler Input Catalogue (KIC) 
is a key tool for selecting stars for the Kepler programme.
The KIC-10 (Kepler Input Catalog, version 10) is placed in the
Multi-mission Archive at the STScI (MAST),%
\footnote{\tt http://\\
archive.stsci.edu/kepler/kepler\_fov/search.php}
with the aim
to provide information useful for the selection of optimum targets for
the mission, not just for the Kepler planet searches, but also for other users,
such as the KASC and the Guest Observer Program.
The KIC reports every known object in a footprint that covers the entire Kepler
field, including the gaps between the Kepler CCDs. It relies on the USBO-B for
the information on faint stars. 
New multiband ground-based photometry in the Sloan g, r, i, and z bands,
plus D51 (a custom intermediate-band filter that includes
the gravity-sensitive Mg b features) was included for essentially 
all the entries in the 2MASS Catalog, and this
photometry has been used to estimate effective temperatures, surface gravities,
metallicities, reddening, and photometric distances.
Masses and radii estimated with the aid of stellar models
are also included in the KIC.

A characteristic precision of the KIC-10 photometry is 0.015 mag in r 
and in the various colours, e.g. g--r, r--i, and i--z.
This is the median error derived from multiple visits to the same stars.
Based on just the photometry, the effective temperatures for solar-type 
stars are uncertain by about 150\,K,
with larger errors towards the hotter and cooler ends of the range covered. 

\section{Organization of the Kepler Asteroseismic Investigation}

%
The goal of the KAI is to ensure fast and efficient utilization of the
unique resource provided by the {\it Kepler} asteroseismic data, as well
as to ensure that these data will remain useful for an extended period after
the end of the mission.

The very large number of stars that are being observed motivates the
involvement of a broad community in the analysis, with direct and flexible
access to the asteroseismic data.
In addition, it is important that the data are analysed in 
a coordinated manner, to optimize the investigations and ensure timely
publication of the results, properly reflecting the underlying efforts.
This is the background for the establishment of the Kepler Asteroseismic
Science Consortium (KASC).
Full access to the asteroseismic data is open to all active members of the
KASC.
A concern for this open data policy, however, has been the possibility of
detection of apparent planetary transits by the KASC members.
Reliable identification of exoplanets with the transit technique requires
extensive analyses beyond the initial detection of a possible transit,
and hence is carefully controlled by the {\it Kepler} Science Team;
similarly, this represents a high-profile goal of the mission, and
hence announcements of planet detections are the responsibility of the
Science PI.
To avoid premature announcements of planet detections all KASC members have
signed a Nondisclosure Agreement, stating that evidence for planet transits
found in the data will be communicated only to the Science Team, for further
analysis.
In addition, the data are checked for obvious exoplanet signatures before
being released to KASC. During the initial release of data from the first quarters (Q0, Q1 and Q2) the short cadence data were modified through a
Transit Removal filter (TRF) aiming at destroying any transit-like features 
in the time series. However, this filter also modified some of the 
oscillation features at low frequencies and it was decided to release 
all {\it Kepler} asteroseismic data without the TRF modification. All KASC data
are now used in a version without filtering (TRF),
including Q0, Q1 and Q2 data.

Information on KASC is available on the KASC web site.%
\footnote{{\tt http://astro.phys.au.dk/KASC/}}
This includes a link to all the relevant documents describing the organization
of the activity.
In particular, the publication strategy and policy 
(DASC/KASOC/0009) should be noted.

\subsection{KASC Working Groups}

%
In order to organize the work among the 400 members of the Kepler Asteroseismic Science Consortium (KASC) a number of working groups (KASC WG) has been created. Each KASC WG has a chair who is responsible for managing and organizing the work related to that WG.

The following WG have been established,
largely according to the relevant types of pulsating stars:

\begin{enumerate}

\item Solar-like p-mode oscillations 
\item Oscillations in clusters 
\item Beta Cephei stars 
\item Delta Scuti stars 
\item Rapidly Oscillating Ap stars 
\item Slowly Pulsating B-stars 
\item Cepheids 
\item Red Giants 
\item Pulsations in binary and multiple stars 
\item Gamma Doradus stars 
\item Compact pulsators 
\item Miras and semiregulars
\item RR Lyrae stars 

\end{enumerate}
Most working groups are further subdivided into subgroups, with responsibility,
e.g., for data analysis, modelling and ground-based support observations.
Working Groups 2 and 9 clearly have a rather special status:
they are intended to deal with those cases where membership of a cluster
or a binary system is relevant for the analysis of a specific {\it Kepler}
asteroseismic target.
In general, it is likely that a KASC member belongs to several working groups
and/or subgroups, depending on his/her interests.
A detailed overview of the working-group structure, including names and
contact information for working-group chairs, is available on the KASC
web site.

Membership of a working group is a condition for access to the 
{\it Kepler} asteroseismic data through KASOC (see below).
To join a working group a KASC member must supply a brief Letter of Intent
to the relevant working group(s) stating his/her interest in the data and plans
for using them.
Access to any {\it Kepler} asteroseismic data is open to all KASC members,
but publication has to be coordinated through the working groups, to ensure
proper credit for the activities.
In the likely not uncommon case that work on a dataset with a specific
purpose uncovers aspects that are relevant to another working group,
the KASC member simply has to join that working group.

The KASC WG are established to ensure an efficient
and structured work within KASC focusing on data analysis, stellar modelling
and publication of data. The chair of each working group (and subgroup)
is responsible for organizing the work within the working group, defining
specific tasks and deadlines and ensuring that all working group members
are involved in the work in agreement with the individual intents expressed in
their KASC Letters of Intent. The WG chairs are also responsible for
internal information within a given working group and for keeping
the KASC Steering Committee informed about the progress of the work
within the working group.
The working groups have very considerable autonomy on how to organize the
work and the publications.

\begin{figure}
\includegraphics[width=80mm]{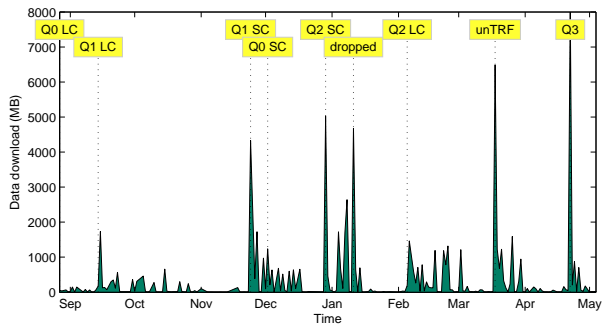}
\caption{Data download as a function of time. The first long cadence
(LC) data were released in September 2009, and at regular intervals we have
released new data to KASC for Q0 -- Q3. Targets dropped by the main science
programme is also included in the KASOC database. In March 2010 we released
all SC data for Q0 -- Q2 without the TRF applied.
}
\label{fig:fig1}
\end{figure}

\subsection{KASOC}

%
To distribute the {\it Kepler} asteroseismic data and help coordinating
the analysis and publications within KASC we have established 
the Kepler Asteroseismic Science Operations Centre (KASOC).%
\footnote{{\tt http://kasoc.phys.au.dk/kasoc/}}
This provides access to the {\it Kepler} time series as well as 
other relevant information about the potential, and actually observed,
{\it Kepler} asteroseismic targets.
It is the intention that the KASOC site will also be used for exchange of
results of the analysis of the {\it Kepler} data, such as modelling, and
ground-based support observations, as well as serving for exchanging
and discussing preprints based on KASC analyses.
Access to KASOC is controlled through an individual password that can be
provided, once the KASC member has joined a working group.
Data downloads are logged.

The SC data are provided to KASOC by Gilliland, after vetting the
time series for possible transits or other conflicts with the {\it Kepler}
exoplanet programme.
The LC data are downloaded directly from MAST%
\footnote{{\tt http://archive.stsci.edu/kepler/}}
(Multimission Archive at Space Telescope Science Institute)

The data are typically made available around 4 months after the end 
of the corresponding quarter.
For example, the data from Q3 (18 September 2009 -- 17 December 2009) were
released 22 April 2010. 
In total 362580 files (May 2010) have been downloaded from
KASOC corresponding to 74 GBytes of uncompressed data (29\,\% of the data
in FITS format and 71\,\% of the download as ASCII datafiles).
Figure~\ref{fig:fig1}
shows the data download as a function of time indicating the peak downloads
in relation to the release of new data.

The KASOC is currently hosted by the Department of Physics and Astronomy.
The intention is that it will be moved to the Royal Library, Copenhagen,
to increase the functionality, including careful version control and
recording of the work processes involved in setting up the data.
Also, by being archived at the Royal Library the long-term preservation and
usefulness of the data can be ensured.

\begin{figure}
\includegraphics[width=80mm]{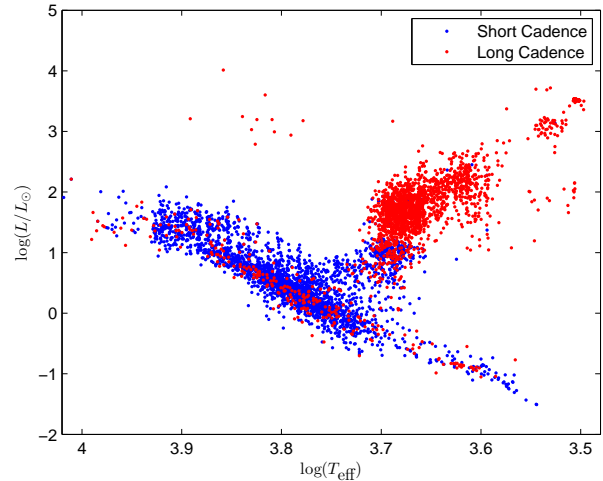}
\caption{The HR diagram for all KASC targets observed during Q0 -- Q3.
SC targets
are mostly main-sequence and dwarf stars (sdB and white dwarfs) while LC
targets are mostly stars with a large stellar radius.
}
\label{fig:fig2}
\end{figure}

\subsection{Survey and specific targets}

%
{\it Kepler} asteroseismology offers unique possibilities for extending
our knowledge about stellar oscillations and stellar structure, and hence
careful planning has been needed to ensure the optimal use of these
possibilities.
This requires a balance between obtaining information about a broad range
of stars and carrying out in-depth investigations of carefully
selected targets.
This is being achieved by dividing the KAI into two, partly overlapping,
phases.
In the initial phase a large number of targets are {\it surveyed}, 
by being observed typically for one month each.
This phase has occupied the first ten months, Q0 -- Q4,
of the mission, leading to the observation of about 5200 targets in SC and
LC.
It has been found that survey of some targets, in particular lower-mass
unevolved solar-like pulsators and compact objects, requires longer periods,
and hence such surveys are being continued in the subsequent quarters,
with observations covering the full quarter.
During Q0 -- Q3 a total of 2937 SC targets were observed for KASC,
spanning the HR diagram.
59 targets were observed for longer than one month and
most of those (44 targets) were observed during the whole mission (Q0 -- Q3).
Based on KIC parameters we show in Figure~\ref{fig:fig2} the 
HR diagram for all KASC targets observed during the Q0 -- Q3 period
(including the LC targets).

On the basis of the surveys, {\it specific} targets have been selected
which will typically be observed for much more extended periods, to
allow detailed studies of the stellar properties, including determination
of aspects of the internal structure and internal rotation
\citep{Christ2010}.
A very interesting aspect, in the case of solar-like oscillations,
is to search for variations of oscillation properties related to
time-varying stellar activity \citep{Karoff2009}.
The phase of observing the specific targets is now starting.

\begin{figure}
\includegraphics[width=80mm]{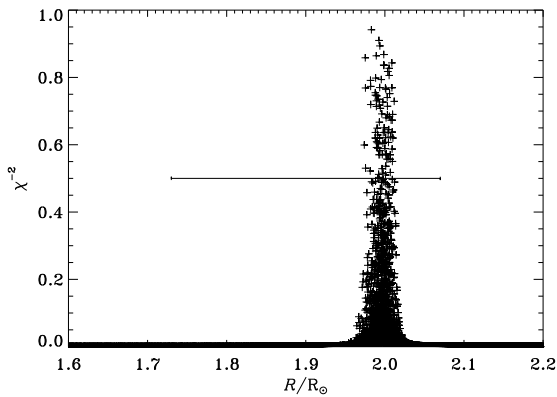}
\caption{Goodness of fit
to observed frequencies based on Q0 and Q1 data for the previously
known planet host HAT-P-7.
Results are shown against radius for a broad range of models
\citep[see][]{Christetal2010}.
The horizontal line illustrates the original determination of the
radius of the star \citep{Pal2008}.
}
\label{fig:hatp7}
\end{figure}

\section{Asteroseismology and exoplanets}

%
There is close synergy between asteroseismology and the study of exoplanets,
providing a strong motivation for the inclusion of the KAI in the
{\it Kepler} mission \citep[see also][]{Kjelds2009}.
Transit observations of an exoplanet provide information about the
ratio between the diameters of the planet and the central star.
Thus an accurate determination of the planetary diameter, of central
importance to the physical characterization of the planet, is totally dependent
on the similarly accurate determination of the radius of the star,
generally far beyond what can be achieved with `classical' astrophysical
observations.
However, from asteroseismic analyses an accurate measure of the stellar
mean density is typically obtained;
fitting the observations to stellar models further provides separate 
and generally fairly precise determinations of the stellar mass and radius.
A radius determination with a precision of better than 3\,\% is typical 
\citep{Stello2009}.
The asteroseismic data also provide information about the evolutionary 
state of the star and hence the age of the star and the planetary system.
This is somewhat dependent on the uncertainties in stellar modelling but
still vastly superior to other types of age determination.
It is an obvious goal of KAI to improve our modelling of stellar evolution
and hence reduce the remaining systematic error in the age determination.

As an example of the potential of asteroseismology for determining
stellar diameters, Figure~\ref{fig:hatp7} illustrates the inference
of the radius of the previously known planet host \hbox{HAT-P-7}
\citep{Pal2008} in the 
{\it Kepler} field, based just on the Q0 and Q1 data \citep{Christetal2010}.
It is evident that, compared with the original determination of the
radius, the precision has been increased by roughly an order of
magnitude, with a corresponding increase in the precision of the determination
of the planetary radius.


In the case of {\it Kepler} most of the central stars of the detected
planetary systems are probably too faint for a detailed asteroseismic
analysis.
In some cases we can expect to be able to determine the mean density of the
star and hence obtain a useful constraint on the properties of the system.
Also, the detailed asteroseismic results obtained for brighter stars
can be used to calibrate the Kepler Input Catalogue (KIC-10)
and hence improve the general precision of stellar parameters.
We note that for the PLATO mission \citep{Catala2009},
under evaluation in the ESA Cosmic Vision programme,
asteroseismic characterization of the central stars in extra-solar
planetary systems is an integral part of the mission.
The results already obtained with {\it Kepler}, illustrated above,
clearly demonstrate the great value of this approach.

\section{Summary and outlook}

%
As pointed out in the sections above asteroseismology is and will become a
very important tool for studying detailed properties of exoplanetary
systems by characterizing the planet-host stars at a level of detail
not possible from other astrophysical techniques. 
This requires improved knowledge of stellar
structure and evolution and hence a general study of 
stellar oscillations in stars
across a wide range of types. The first results
of the asteroseismic analyses from Kepler were described by \citet{Gillil2010a}.
These already showed the huge potential of Kepler for
asteroseismology, including high-quality asteroseismology of a few
solar-like stars \citep{Chapli2010},
and analysis of a large number of red giants
showing that high signal-to-noise photometry 
can provide direct detection of oscillation power for hundreds of stars,
bringing red giant seismology a huge step forward
\citep{Beddin2010}.
Also, as discussed by \citet{Gillil2010a} important results have been 
obtained for classical pulsating stars.
The main improvement
for those stars is the high stability of the Kepler photometer and the
extended length of the time series, providing a great improvement in
frequency accuracy of the detected modes and low-amplitude
detectability \citep{Kolenb2010}.
An interesting feature 
is the large number of hybrid stars where one star
shows simultaneous oscillations of different types, e.g.,
delta Scuti and gamma Doradus pulsations \citep{Grigah2010}.
Just these early results 
are orders of magnitude better than seen before and
this indicates how the future analysis of Kepler data for many types of
stars will impact our general understanding of stellar structure
and evolution. With Kepler going from the survey phase to the
specific target phase (with stars being observed for extended periods of time)
we will increase the frequency accuracy to a level where changes in the
stellar structure may be observed for main-sequence stars.
The KAI and KASC will play a key role in the 
continuing use and analysis of the Kepler asteroseismic data.
%
%
%
%

\acknowledgements
Funding for the {\it Kepler mission} is provided by NASA's 
Science Mission Directorate.
We thank the entire {\it Kepler} team for the development and operations
of this outstanding mission.
This work was supported by the European Helio- and Asteroseismology
Network (HELAS), a major international collaboration funded by the
European Commission's Sixth Framework Programme.

\newpage


\end{document}